# Incentivizing the Dissemination of Truth Versus Fake News in Social Networks

Abbas Ehsanfar
School of Systems and Enterprises
Stevens Institute of Tech., Hoboken, NJ
aehsanfa@stevens.edu

Mo Mansouri
School of Systems & Enterprises
Stevens Institute of Tech., Hoboken, NJ

*Abstract*—The concept of "truth," as a public good is the production of a collective understanding, which emerges from a complex network of social interactions. The recent impact of social networks on shaping the perception of truth in political arena shows how such perception is corroborated and established by the online users, collectively. However, investigative journalism for discovering truth is a costly option, given the vast spectrum of online information. In some cases, both journalist and online users choose not to investigate the authenticity of the news they receive, because they assume other actors of the network had carried the cost of validation. Therefore, the new phenomenon of "fake news" has emerged within the context of social networks. The online social networks, similarly to System of Systems, cause emergent properties, which makes authentication processes difficult, given availability of multiple sources. In this study, we show how this conflict can be modeled as a volunteer's dilemma. We also show how the public contribution through news subscription (shared rewards) can impact the dominance of truth over fake news in the network.

I. INTRODUCTION

In today's social network era, the percepton of truth depends on emerging social interactions along with the traditional news broadcasters. These online social interactions leaves us with a diverse pool of news resources most of which are all but impossible to be individually fact checked and verified in terms of credibility and truthfullness. The lightening spread of news in the social media is known to be one aspect of network diffusion in social networks [1]. The network effect in diseminating news suggests that a piece of news gains credibility when goes viral. In other words, higher number of redistribution for a piece of news by other nodes is often perceived as its authenticity accross a networked system [2] [3] [4] [5]. Activity of these nodes (agents) also depends on other agents in complex network of social interactions. At the same time, accessing the stream of true and credible news is costly. However, creating such credible streams in social networks to eliminate dominancy of fake news will be ultimately beneficial for the whole society, in long run. This suggests that the truthfull news (i.e. truth) can be considered as a public good.

In networked social systems with public goods, public service has cost for volunteering agents while it generates benefit that belongs to all members [6]. In social networks, disemminating truth is considered a public good as costly process of attaining the truth and the public benefit of the news disintencivize the social members to volunteer in the public process and instead incentivize them to rely on other agents in the networked system.

Public good game in social systems is a model of situation where the presense of free-rider incentives jeopardizes the socially beneficial outcome [7]. Accordingly, in the volunteer's dilemma in social systems, higher number of social members results in lower level of active participation in validating and authenticating the news (public goods). This conflict towards the volunteering equilibrium is raised in many forms: cooperation among the members [8] [9], inclusion of heterogeounous agents among the players [10], which is often refered to as as "strong" and "weak" members [9] [11], punishing defectors in the system [7] [12] [13], rewarding the volunteers using shared reward pool [14] [15], and so forth. The incentivising mechanism for such circumstances is among the well studied areas and multiple studies have shown the mathematical merits of it, including the existance and stability of its equilibrium points in evolutionary games [13] [16]. The incentivizing mechanim and the principle disemination, as two basic ideas behind this study, are two pillars of governance in systems of systems [17].

Game theoretical approaches are applied to model the equilibrium in social systems with heterogeneous groups of agents. The attacker-defender game is applied to protect power system against intelligent external attacks [18]. A two-step game-based formulation can optimize the cloud service request by distributed mobile devices [19]. A game theoretic payment sharing model could incentivize the truthfull participation of agents in energy market [20] [21]. Finally, a mathematical framework that incentivizes the agent movements also reduces the collective cost of a transportation network [22].

In this study, we introduce a mathematical integrated framework, based on game theoretic mixed-strategy formulation of volunteer's dilemma, to model disemination of truth and fake news in social networks.

*A. Basic Model Assumtions*

In this study we introduce two types of agents within the population: regular agents and the fake news agents. We assume the former population is greater than the latter in size (10-fold) when the collective benefit of disemination of truth (public good) belongs to all players and the group benefit of disemmination of fake news merely belongs to the group of fake news agents. Also, truth is assumed to be outcome of social intereaction and dominance between the regular and fake news groups. We assume that the agents are distriubted and network is balanced among agents (either strategically or randomly), which implies that the group with higher number of volunteers can expect the dominance of its message in the social network. For rewardign mechanism, we assume that the subscription based journalism works as a rewarding mechanims and volunteering agents receive more or less the same share of the reward. We also assume that the effects of fake news is not directly seen by the regular agents, i.e. regular agents cannot categorically and directly punish fake news agents. An implication of the latter assumption is that the equilibrium of fake news model doesn't affect the equilibrium of regular news. The opposite statement is not true. Accordingly, we formulate the probabilistic model of fake news based on the equilibrium of probabilitic model of regular users. Finally, we assume that the users among each group of agents (regular and fake news agents) are homogenous, then the model is symetric for each group.

In the following section, the volunteer's dilemma is illustrated, then the modified formulation of symetric volunteer's dilemma (VOD) is expressed, and the payoff formulations of fake news agents are presented. In the third section, we show the analytical equilibrium of volunteering model defined as the net payoff versus the volunteering ratio. The effect of shared reward and group size is shown on the equilibrium points. Lastly, we caluculate the equlibium points for disseminating the fake news based on the model equilibrium of regular users.

## II. MODEL

To start with, we introduce the model variables and parameters. A number of regular agents (N) are interacting in social system. These regular agents can either participate (M) in validating and authenticating the truthful news or consume news (N-M). The minimum group size of volunteers is required to achieve the public good (k). The cost of volunteering (c) and the cost of failing to achieve the public good ($\alpha$) occurs in case. Having an rewarding mechanism, $\sigma$ is distributed among the volunteering agents. Because the model is a symetric model, the equilibrium shows the volunteering ratio or the probability of volunteering among users (p). On the other side of aisle, F number of fake news agents with cost $c_f$ volunteer in publishing fake news. Both are fraction of N and c respectively. In the rest of this paper, we refer to act of not volunteering (sharing/distributing but not authenicating and validating) as defecting. We formulate the effective features for agent behavior in social networks and express the model equations next. Table 1 shows the variables, mathematical symbols, and their description.

*Table 1. Variables and Description*

| Variable | Table column subhead |
|---|---|
| N | Number of regular users that could potentially disemminate truth in the network |
| M | Number of volunteering agents among regular agents |
| c | Cost of volunteering for regular agents |
| α | Cost of failing to dominate the news cycle |
| k | Minimum group size to dominate the news cycle: the number of volunteering fake news agents |
| x | Volunteering ratio and the probability of volunteering in symetric model |
| σ | Aggregated reward shared among regular agents |
| $p_v(M)$ / $p_v(k)$ | Payoff of volunteering for a regular/fake news agent as function of the number of volunteers |
| $p_d(M)$ / $p_d(k)$ | Payoff of defecting for a regular/fake news agent as function of the number of volunteers |
| $\overline{P_v}(k)$ / $\overline{P_d}(k)$ | Average payoff of volunteering/defecting regular agents as function of group size |
| F | Number of fake news users that could potentially diseminate fake news |
| $c_f$ | Cost of volunteering for fake news agents |
| $\overline{P_v}(M)$ / $\overline{P_d}(M)$ | Average payoff of volunteering/defecting fake news agents as a function of regular volunteering agents |
| P* | Equilibrium probability of volunteering for regular agents |
| $\overline{P_v}$ / $\overline{P_d}$ | Expected value of volunteering/defecting payoff of fake news agents |

*A. Volunteers's Dilemma*

Equation 1 shows the individual payoff in volunteer's dilemma (VOD) for N players. At each round, N random social network users, among infinite number of people, and potential volunteers among them are selected. The players produce the public good if at least k players decide to volunteer (1≤k≤N). The cost of volunteering is c, relative the baseline payoff of 1, and the cost is mandatory whether or not the public good is achieved. If the number of volunteers is less than k, the failing cost of α>c occurs for each player.

$$p_v(M) = \begin{cases} 1-c \\ 1-a-c \end{cases} \quad p_d(M) = \begin{cases} 1 & if\ M \geq k \\ 1-a & if\ M < k \end{cases} \quad (1)$$

where M is number of volunteers and k is minimum number of volunteers to achieve public good.

*B. Payoff Model of Truth Dissemination*

In the general VOD with shared reward among the volunteers, the average payoff for volunteering is formulated by modifying the reward sharing model of generalized volunteer's dilemmas (Chen, 2013). To express the equations, we unify the terminilogy of variables in the model with the introduced variables in Table 1. Accordingly, number of players represents regular users and number of volunteers represents users volunteering in disseminating truth, etc.. Eq. 2 shows the average payoff of volunteering agents having a fixed shared reward distributed among them:

$$\overline{P_v}(k) = \sum_{M=k-1}^{N-1} \binom{N-1}{M} x^M (1-x)^{N-M-1}(1-c)$$
$$+ [1 - \sum_{M=k-1}^{N-1} \binom{N-1}{M} x^M (1-x)^{N-M-1}](1-c-a)$$
$$+ \sum_{M=k-1}^{N-1} \binom{N-1}{M} x^M (1-x)^{N-M-1}(\frac{\sigma}{M+1} - \frac{\sigma}{N})$$
(2).

The average payoff of regular users when they don't participate in diseminating the truthful news is expressed as:

$$\overline{P_d}(k) = \sum_{M=k}^{N-1} \binom{N-1}{M} x^M (1-x)^{N-M-1}$$
$$+ [1 - \sum_{M=k}^{N-1} \binom{N-1}{M} x^M (1-x)^{N-M-1}](1-a)$$
$$- \frac{\sigma}{N} * \sum_{M=k}^{N-1} \binom{N-1}{M} x^M (1-x)^{N-M-1}$$
(3)

where σ is shared reward among volunteers and x is the ratio of volunteers or volunteering probability.

### C. Payoff Model of Fake News Dissemination

In the previous section, we modified a model of VOD for regular social media users. We also introduce another model of VOD with different formulation for users who are incentivized to disseminate fake news. While payoff mechanism for the former group of users is formulated in Eq. 1, the payoff model of latter groups of users include different parameters and excludes the rewarding mechanism. Here, the intuition is the fake news agents cannot benefit from dominance of truth in social network and don't contribute to disemination of truth through shared rewardign mechanism. Eq. 4 shows the individual payoff of fake news agents and Eq. 5 and 6 show the average payoff of volunteering and defecting to contribute to fake news community respectively:

$$p_v(k) = \begin{cases} 1-cf \\ 1-a-cf \end{cases} \quad p_d(k) = \begin{cases} 1 & if\ k \geq M \\ 1-a & if\ k < M \end{cases}$$
(4)

$$\overline{P_v}(M) = \sum_{k=M-1}^{F-1} \binom{F-1}{k} x^k (1-x)^{F-k-1}(1-cf)$$
$$+ [1 - \sum_{k=M-1}^{F-1} \binom{F-1}{k} x^k (1-x)^{F-k-1}](1-cf-a)$$
(5)

$$\overline{P_d}(M) = \sum_{k=M}^{F-1} \binom{F-1}{k} x^k (1-x)^{F-k-1}$$
$$+ [1 - \sum_{k=M}^{F-1} \binom{F-1}{k} x^k (1-x)^{F-k-1}](1-a)$$
(6)

### D. Integrated Incentivizing Model

In social network, we combine the potentially truthful and fake news agents. The former set of users share the collective reward of disseminating truth in the network. The second set of agents could achieve in-group benefits of spreading fake news when disseminating truth affects their utility through the cost of failure. In the combination of these agents in network, the smaller and less public group, i.e. fake news agents, calculate their expected payoff as a function of expected number of volunteers among regular agents. The volunteering ratio being estimated from the Eq. 2 and 3 equilibrium, fake news agents infer a bionomial distribution for the number of volunteering agents. Eq. 7 shows the probabilistic sum of averaged payoff for fake news agents using probability mass function (pmf) of volunteering agents:

$$\overline{P_v} = \sum_{M=0}^{F} \binom{N}{M} p^{*M}(1-p^*)^{F-M} \overline{P_v}(M)$$

$$\overline{P_d} = \sum_{M=0}^{F} \binom{N}{M} p^{*M}(1-p^*)^{F-M} \overline{P_d}(M)$$
(7)

where $v$ represents the volunteering and $d$ represents the defecting group. This expected payoff shows the potential incentive for fake news agents to actively participate in disseminating fake news in the social network. In effect, this model shows the effects of rewarding mechanism on incentivizing the dominance of truth in the network. In the next section, we explore the analytical effect of rewarding mechanism and model parameters on symetric equilibriium points for disemminating truthful and fake news.

### III. RESULTS

To show the analytical results, we first review our numeric assumptions for parameters of the integrated model. We assume one hundred regular users in the social network. Arround one twelfth of this number are assumed as potential fake news agents (8). We create a fictional currency and represent it by ç. We pick ç0.5 for volunteering cost (c) and ç0.9 for failing cost (α). As it was mentioned before, the cost of creating fake news is a fraction of cost of volunteering for disseminating truth (cf = ç0.1). For the rewarding mechanism, we assume a shared reward between ç5 and ç8, which is evenly distributed among the volunteering agents. For the minimum volunteering group size, based on the maximum number of active fake news sources, we assume a range of numbers less than 8.

To run the model and show the results, we define two comparing factors (Table 2). The first factor calculates the difference between averaged payoff of volunteering and defecting agents. Apparently, the measure shows whether the agents benefit from either volunteering or defecting depending on the difference being positive or negative. For the second factor, we consider the fake news agents and based on Eq. 2 and 3, we calculate the expected difference between the payoff of volunteering and defecting (Eq. 7) among fake news agents.

Table 2.    COMPARING FACTORS

| Factor | Description |
|---|---|
| $\overline{P_v}(k) - \overline{P_d}(k)$ | Net difference between the average payoff of volunteering and defecting (regular agents). Net value equal to zero shows a potential equilibrium for volunteering ratio among regular agents. (Fig. 1 and 2) |
| $\overline{P_v} - \overline{P_d}$ | Net difference between the expeccted payoff of volunteering and defecting among fake news agents, which shows whether the "fake news" agents are incentivized to volunteer or defect. Net value equal to zero indicates a potential equilibrium for volunteering ratio among the fake news agents. (Fig. 3) |

To explore and discuss the analytical results, first we show the possible equilibrium points for volunteering ratio among the regular agents. We approach the equilibrium points from two perspective: the effect of shared reward on the expected payoff of volunteering and the that of minimum group size on both metrics. Then, we discuss the effect of model equlibriums on the expected payoff of volunteering among fake news agents.

Increasing the shared reward among the volunteering agents significanly affects the volunteering payoff and the potential equlibrium points. Fig. 1 shows the net payoff corresponding to the volunteering probability for a range of total shared rewards. In this case, the group size is selected equal to 6, which represents minimum number of volunteering agents to achieve the public good. In this figure, the net payoff equal to zero creates two sets of equilibrium points. Both sets show equlibrium because for the net payoff of zero, agents don't have incentive to either increase or decrease their volunteering probability. However, one set of those equilibrium points are stable while the other set are unstable points. Intuitively, without proving the stability by the mathematical expressions, the set with positive derivative is unstable because any change in volunteering ratio reinforces further change to the same direction (i.e. increasing ratio results in higher payoff for volunteering that incentivizez higher ratio). Hence, the second set (with negative derivative) shows the stable equlibrium points for volunteering ratio. This figure shows that higher amount of shared reward leads to higher stable equlibrium points in the symetric model. We know that the dominant strategy for low values of shared reward (σ<4) is defecting (Eq. 3 > Eq. 2).

In the second case, we show the effect of minimum group size on net payoff and equlibrium points. We select the range of 5 to 8 for group size (lower than the number of potential fake news agents) and 5 for shared reward. Fig 2 shows that the group size has considerable effect on the first set of equilibrium points but negligible effect on the second set (stable). A reason is low variance of binomial distribution because of relatively high number of regular agents. This means that certain probability either pass the group threshold or doesn't and it is very improbable to fall in between. Accordingly, the higher the group size is, the lower the first equilibrium will be. For high number of group size (>8), the dominant individual strategy is defecting.

In our model, dissemination of fake news happens when the number of fake naws sources dominates the number of truthful sources (k>M in Eq. 5 and 6). The fake news agents might be incentivized to volenteer knowing about the volunteering preference among regular users. In this study, we claim that the rewarding mechanism for regular users could be designed as disincentivizing mechanism for participation among fake news agents. Lower volunteering among fake news agents means higher probability of disemination of truth in the network. Having these being said, we implement the Eq. 7 using the equilibrium points for regular users (p* in Eq. 2 and 3). In Fig. 3, we show the net payoff for volunteering and defecting among the fake news agents. The selected range of volunteering equilibrium points (legends) shows the realistic points calculated using the first model (Fig. 1 and 2). The expected volunteering ratio has significant effect on the fake news equlibrium points. For lower volunteering points, the net payoff of participation in disseminating fake news increases and the equilibrium points shift toward the lower values, which means that relatively lower level of activity among fake news agent can incentivize those agents to remain active. Another effect of volunteering ratio is seen on the amount of net payoff for participation among the fake news agents. The potential net payoff is significantly higher when the volunteering ratio is expected to be lower, i.g. 4 percent volunteering probability shows maximum net payoff 4-fold of the case with 10 percent probability. Finally, higher volunteering ratio shifts the first set of equilibrium points (unstable) to right. Although these points are unstable, its implication on adopted strategy by fake news agents is noticble: any participation rate lower than those points could incentivize lowering the participation rate to zero

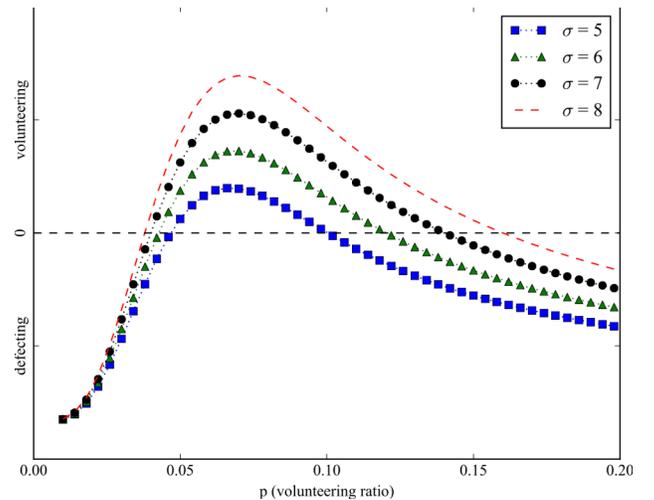

Figure 1. Net payoff difference for volunteering and defecting agents corresponding to users' volunteering ratio. The σ shows total reward (cost) distributed (shared) among volunteers (users).

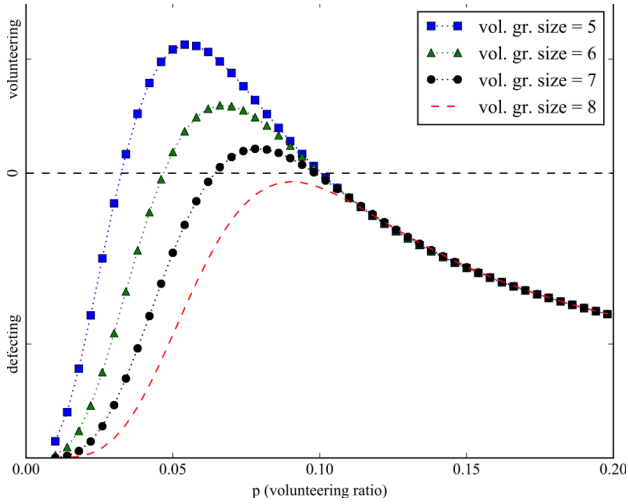
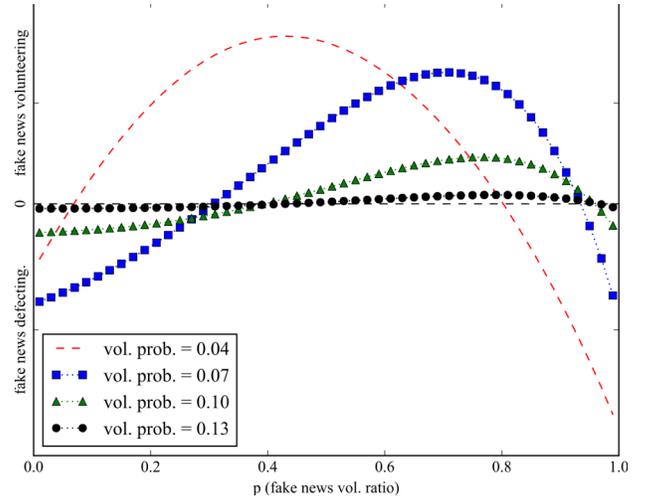

Figure 2. Net payoff difference for volunteering and defecting agents corresponding to users' volunteering ratio. The "vol.gr.size" is the minimum group size for volunteers to produce public good.

Figure 3. Net payoff for volunteering and defecting among "fake news" agents, corresponding to fake users' volunteering ratio. The "vol.prob." is the volunteer's equilibrium probability for the regular agents.

(negative derivative of net payoff).

Last but not least, the reverse effect of equlibrium points among fake news agents on the main model should be discussed. The participation rate among the fake news agents affects the volunteering model of regular users through group size (k). Increasing the number of participants among fake news agents also increases the minimum group size to achieve the public good by the regular users. However, as was discussed before, the effect of the group size on the stable equilibrium points in Fig. 2 is minimal, i.e. the equilibrium point of volunteering remains around 0.09 for range of feasible group size (<8). This imply that the regular users are minimaly incentivized by equilibrium behavior of fake news agents.

## IV. Discussion and Conclusion

In sum, the effect of real-world financial behavior among social network users, such as subscription to credible journalism, on disemination of fake news can be discussed using analytical game theoretic models. In this paper, we proposed an integrated model combinig two VOD models of volunteering among regular users and fake news agents in social network. The integrated model attacks the systemic problem of emerging fake news against the fact-based news (truth) in social networks, which manifests a complex System of Systems. To show the impact of agent adopted policies on spreadng fake news in networked social systems, we formulated a rewarding mechanism that incentivizes volunteering in the system toward dissemination of truth. The equlibrium points represent the agent adopted policy in a symetric system of systems. This particular mechanism (agent adopted policy) uses a shared subscription program with a fixed subscription fee for regular agents (users). We formulated a payoff mechanism for fake news agents. Integrating the volunteering model with the fake news participation model, we observed the effects of a shared rewarding mechanism on the equlibrium points of activity among fake news agents. We showed that relatively marginal variation in shared reward has significant effect on the disemmination and ultimate dominance of fake news in the network. We discussed that the equlibrium points for fake news agents heavily depends on the incentivized system equilibrium while the reverse effect is not observable.

For future studies, the mathematical illustration of the integrated equlibriums helps to establish a mechanism which incentivizes the dominance of truth in social networks. In addition, considering asymetric models with heteregenous agents can considerably improve the explanatory aspect of our proposed model. Finally, the dynamic algorithms can be applicable to design rewarding mechanims that target collective benefit of the System of Systems.